\documentclass[usenatbib]{mnras}

\usepackage{newtxtext,newtxmath}
\usepackage[T1]{fontenc}
\usepackage[utf8]{inputenc}
\usepackage{ae,aecompl}
\usepackage[dvipsnames]{xcolor}
\usepackage{cancel}

\usepackage{graphicx}	
\usepackage{amsmath}	
\usepackage{amssymb}	
\usepackage{pifont}

\newcommand{\Ms}{\ensuremath{\, M_s\, }}
\newcommand{\pc}{\,\mathrm{ pc }}
\newcommand{\Mdot}{\,\mathrm{ M}_{\odot} }
\newcommand{\Msun}{\,\mathrm{ M}_{\odot} }
\newcommand{\kpc}{\,\mathrm{ kpc }}
\newcommand{\Mpc}{\,\mathrm{ Mpc }}
\newcommand{\Myr}{\,\mathrm{ Myr }}
\newcommand{\Gyr}{\,\mathrm{ Gyr }}
\newcommand{\yr}{\,\mathrm{ yr }}

\newcommand{\km}{\,\mathrm{ km }}

\newcommand{\cm}{\,\mathrm{ cm }}
\newcommand{\s}{\,\mathrm{ s }}
\newcommand{\G}{\mathrm{G}}

\newcommand{\erg}{\,\mathrm{ erg }}
\newcommand{\K}{\,\mathrm{ K }}
\newcommand{\amu}{\,\mathrm{ amu }}

\newcommand{\vmark}{\ding{51}}
\newcommand{\xmark}{\ding{55}}

\newcommand{\Ramses}{\texttt{Ramses}}
\newcommand{\ie}{\emph{i.e.}\, }

\newcommand{\hugo}[1]{#1}

\title[Lonely high z BHs]{The erratic dynamical life of black hole seeds in high-redshift galaxies}

\author[H. Pfister et al.]{
Hugo Pfister,$^{1}$\thanks{E-mail: pfister@iap.fr}
Marta Volonteri,$^{1}$
Yohan Dubois,$^{1}$
Massimo Dotti$^{2,3}$
and
Monica Colpi$^{2,3}$
\\
$^{1}$Sorbonne Universit\'{e}s, UPMC Universit\'{e} Paris 06 et CNRS, UMR7095, Institut d'Astrophysique de Paris,\\
98bis boulevard Arago, F-75014, Paris, France\\
$^{2}$Dipartimento di Fisica G. Occhialini, Universit$\grave{a}$ degli Studi di Milano--Bicocca, Piazza della Scienza 3, I-20126 Milano, Italy\\
$^{3}$INFN, Sezione Milano--Bicocca, Piazza della Scienza 3, I-20126 Milano, Italy
}

\date{Accepted XXX. Received YYY; in original form ZZZ}

\pubyear{2017}

\begin{document}
\label{firstpage}
\pagerange{\pageref{firstpage}--\pageref{lastpage}}
\maketitle

\begin{abstract}
The dynamics of black hole seeds in high redshift galaxies is key to understand their ability to grow via accretion and to pair in close binaries during galactic mergers.
To properly follow the dynamics of black holes we develop a physically motivated model to capture unresolved dynamical friction from stars, dark matter and gas.
We first validate the model and then we use it to investigate the dynamics of seed black holes born at $z\sim9$ in dwarf proto-galaxies.  We perform a suite of zoom cosmological simulations with spatial resolution as high as 10~pc and with a stellar and dark matter mass resolution of $2\times10^3 \,\Msun$ and $2\times10^5 \,\Msun$ respectively. We first explore the dynamics of a seed black hole in the galaxy where it is born and show that it is highly erratic if the seed mass is less than  $10^5\, \Msun$. The dynamics is dominated by the stellar component, whose distribution is irregular and patchy, thus inducing stochasticity in the orbits: the black hole may be anywhere in the proto-galaxy. When this dwarf merges into a larger galaxy, it is paramount to simulate the process with very high spatial and mass resolution in order to correctly account for the stripping of the stellar envelope of the satellite black hole. The outcome of the encounter could be either a tight binary or, at least temporary, a wandering black hole, leading to multiple black holes in a galaxy, each inherited from a different merger.

\end{abstract}


\section{Introduction}
\label{Introduction}

The high redshift Universe is the birthplace of the seeds of the supermassive black holes (BHs) observed in today's galaxy center \citep{KormendyHo2013}. A variety of different physical mechanisms for seed formation have been proposed \citep[][and references therein]{2018arXiv181012310W}, but observational constraints are hampered, since the seeds are predicted to have relatively low masses ($10^2-10^5 \, \Msun$) and form at high redshift ($z>6$), making their electromagnetic emission faint \citep{2016PASA...33...54R}. 

The seeds build up their mass via accretion of gas and stars, or via mergers with other BHs 
\citep[e.g.][]{Volonteri_03}.
When BHs merge, they emit gravitational waves, and detection of such waves provides a complementary way of probing BH seeds \citep{2007MNRAS.377.1711S,2012MNRAS.423.2533B,2018MNRAS.481.3278R,2018MNRAS.479L..23H,2018arXiv181011033D}. For BHs with masses in the range $10^4-10^7\, \Msun$ the gravitational waves have frequency around mHz, and they are therefore primary targets for LISA, which can detect BHs with such masses out to $z>20$ \citep{LISA_Proposal}.

However, before coalescing  by emission of gravitational waves, which can merge BHs of $10^4-10^7\, \Msun$ in less than a Hubble time once their separation is $\sim10^{-4}-10^{-2}$~pc,  BHs have a long journey 
\citep{Begelman_80}.
They are initially separated by tens of kpc and sit in the center of separate galaxies, which eventually merge. Then, the long process of dynamical friction \citep{Chandrasekhar_43} begins, driving BHs toward the center of the galaxy remnant, until they form a binary when their separation is pc-scale 
\citep[e.g.][and references therein]{Mayer2007,Pfister_17}.
Once the binary has formed, scattering with stars 
\citep[e.g.][]{1996NewA....1...35Q,2007ApJ...660..546S,Khan2012,Vasiliev2015}
, interactions with massive or circumbinary discs \citep[e.g.][]{Dotti2007,2009ApJ...700.1952H,Goicovic_17} or even three-body scattering with another incoming BH
\citep[see][and references therein]{Bonetti18-triplets} 
are invoked to bridge the final gap to where emission of gravitational waves becomes efficient.

Cosmological simulations are excellent tools to study the properties of BH evolution over cosmic time, since they can track the joint evolution of BHs and of the galaxies they are embedded in \citep{Tremmel_18}. Large-volume simulations provide good statistics, having a large number of galaxies and BHs in their boxes, but lack of mass and spatial resolution means that not even the formation of BH binaries can be resolved. Zoom simulations can have much higher resolution, but they allow for the study of a limited number of galaxies and BHs. In this paper, we present a model to better track the dynamics of BHs, validate it and show its limitations. We then use our model in high-resolution zoomed cosmological simulations 
to study the yet unexplored dynamics of BHs
of mass $10^4\Msun-10^5\Msun,$ in a cosmological context,  primary targets for the LISA observatory \citep{LISA_Proposal} .

\section{Dynamical friction in numerical simulations}
\label{Subgrid_model_for_dynamical_friction}
Due to their high mass, BHs attract surrounding material, gas, stars and dark matter, which create an overdensity lagging their passage. This overdensity drags and decelerates the moving BH: this phenomenon  is referred to as dynamical friction \citep{Chandrasekhar_43,Chapon_13}.
To resolve the resulting force in a numerical simulation, \cite{Pfister_17} have shown that the spatial resolution, or the softening, should be smaller than the influence radius
\begin{equation}
r_\text{inf} = \frac{\G M_\bullet}{\sigma^2} =1 \pc \left( \frac{M_\bullet}{10^7 \Mdot} \right) \left( \frac{\sigma}{200\km \s^{-1}} \right)^{-2} \,, \label{r_inf}
\end{equation}
where $M_\bullet$ is the mass of the BH and $\sigma$ is the velocity dispersion of material (gas, stars or dark matter) around the BH. This is because the typical size of the drag, partly causing dynamical friction, has a typical size of the same order as $r_\text{inf}$ \citep{Colpi_99}.
In cosmological simulations, the typical resolution is $\sim 100 \pc-1 \kpc$, much larger than the pc-scale needed to resolve $r_\text{inf}$ for a $10^7 \Mdot$ BH in a Milky-Way like galaxy. Therefore, we must remove by hand the momentum that a BH would lose through dynamical friction if we were able to resolve the phenomenon.
In this section we first describe how we implement unresolved dynamical friction in the adaptive mesh refinement code \Ramses~ \citep{Teyssier_02} for collisionless particles (stars and dark matter); the code already includes a correction for dynamical friction from gas \citep{Dubois_12}.

We follow an approach similar  to \cite{Tremmel_15}, although we include not only the contribution to dynamical friction from slow moving particles but also from fast moving particles, which can play an important role when the density profile becomes shallow \citep{Antonini_12a,Dosopoulou_17}.

We measure all the quantities needed to estimate dynamical friction in a sphere $\mathcal{S}$ centered on the BH with a radius 4$\Delta x$, where $\Delta x$ is corresponds to the minimum grid size. We chose $\mathcal{S}$ to be consistent with the already existing implementation for gas accretion, feedback and dynamical friction \citep{Dubois_12}.

We report here Eq.~(30) from \cite{Chandrasekhar_43}. This gives an analytical estimate of the amount of momentum that must be removed to BHs due to dynamical friction: 

\begin{eqnarray}
\vec{a}_\text{DF} = -4 \pi \G ^2 M_\bullet \frac{\vec{v_\bullet}}{v_\bullet ^3}  (
\text{ln} \Lambda  \int_0^{v_\bullet} 4\pi v^2 f(v) d v 
 + ...   \nonumber\\
 ...  \int_{v_\bullet}^{\infty} 4 \pi v^2 f(v) \left[ \text{ln} \left( \frac{v+v_\bullet}{v-v_\bullet} \right) - 2\frac{v_\bullet}{v} \right] d v ),  
 \label{dynamical friction}
\end{eqnarray}
where we denote as $M_\bullet$ the mass of the BH, as $\vec{v_\bullet}$ (with magnitude $v_\bullet$) the relative velocity of the BH with respect to the velocity of the background, and $\tilde{\vec{v}}$ defined below in Eq. (\ref{vtilde});
$\text{ln}\Lambda=\text{ln}(4\Delta x/r_\text{def})$ is the Coulomb logarithm (this expression is justified below); and $f$ is the distribution function:

\begin{equation}
4\pi v^2 f(v) = \frac{3}{256\pi \Delta x^3} \sum_{i\in \mathcal{S}} m_i \delta (v_i - v).
\end{equation}
Here $\vec{v_i}$ (with magnitude $v_i$) is the relative velocity of particle $i$ with respect to the velocity of the background, $m_i$ is the mass of particle $i$ and $\delta$ is the Dirac function.

The velocity of the background, $\tilde{\vec{v}}$, is simply the mass-weighted velocity of all particles (except the BH particle) enclosed in $\mathcal{S}$:
\begin{equation}
\tilde{\vec{v}}=\frac{1}{M}\sum_{i\in \mathcal{S}}\vec{v_i} m_i \, ,
\label{vtilde}
\end{equation}
where $M$ is the total mass enclosed in $\mathcal{S}$. We stress here that the background velocity is computed for stars and dark matter separately, the reason is that dynamical friction assumes particles with similar masses, which is a reasonable assumptions if we consider an assembly of stars, and an assembly of dark matter particles, but not if we consider stars and dark matter particles together. Therefore we compute the contribution from dark matter, $\vec{a}_{\rm DF,DM}$, and stars, $\vec{a}_{\text{DF},\star}$ separately.

We justify here the expression above for the Coulomb logarithm $\text{ln} \Lambda=\text{ln}(4\Delta x/r_\text{def})$.  In the classical derivation of dynamical friction \citep{Chandrasekhar_43}
the Coulomb logarithm represents the ratio between the ``minimum'' and ``maximum'' impact parameters that affect the velocity change. The minimum impact parameter represents that required to have a deflection of 90$^\circ$, which in the Keplerian case is \hugo{the deflection radius}:
\begin{eqnarray}
r_\text{def} &=& {\G M_\bullet}/{v_\bullet^2} \label{eq:InfluenceRadius}\\
&\simeq& r_\text{inf}  \,
\end{eqnarray}
while the maximum impact parameter is the distance at which the stellar density becomes  sufficiently ``smaller'' than around the BH to become insignificant in modifying its velocity. 
In our case, gravity is computed self-consistently by the code outside $\mathcal{S}$; therefore, the integration must be stopped at $4\Delta x$ if we do not want to double count dynamical friction. This naturally leads to $\text{ln}\Lambda=\text{ln}({4\Delta x}/{r_\text{def}})$. Furthermore, as explained in \cite{Beckmann_17}, using subgrid models when resolution is sufficient to account for dynamical friction can lead to incorrect results. For this reason, when $4\Delta x \leq r_\text{def}$, we set $\vec{a}_\text{DF}$ to 0.

\section{Additional physics: galaxies and black holes}
\label{section:SubgridPhysics}

\texttt{Ramses} 
follows the evolution of the gas using the second-order MUSCL-Hancock scheme for the Euler equations. The approximate Harten-Lax-Van Leer Contact \citep{Toro_97} Riemann solver with a MinMod total variation diminishing scheme to reconstruct the interpolated variables from their cell-centered values is used to compute the unsplit Godunov fluxes at cell interfaces. An equation of state of perfect gas composed of monoatomic particles with adiabatic index $\gamma=5/3$ is assumed to close the full set of fluid equations. Collisionless particles (dark matter, stellar and BH particles) are evolved using a particle-mesh solver with a cloud-in-cell interpolation. The size of the cloud-in-cell interpolation is that of the local cell for BHs and stars, however, dark matter particles can only project their mass on the grid down to a minimum cell size of $\Delta x_{\rm DM}>\Delta x$ (as these particles are usually larger in mass than stars or gas, we smooth their mass distribution to reduce their contribution to shot noise). When cloud-in-cell interpolation is used, therefore, even if the mass of the dark matter particle is larger than the BH mass, since the dark matter distribution is smoothed, scattering off dark matter particles becomes unimportant.

Gas is allowed to cool by hydrogen and helium with a contribution from metals using cooling curves from \cite{Sutherland_93} for temperatures above $10^4 \K$. For gas below $10^4 \K$ and down to our minimum temperature of $10 \K$, we use the fitting functions of~\cite{Rosen_95}.

Star formation is stochastically sampled from a random Poisson distribution \citep{Rasera_06}: at each timestep $\Delta t$, in each cell of size $\Delta x$ containing a gas mass $M_\mathrm{gas}$, the mass of newly formed stars, $M_{\star,\mathrm{new}}$, follows a Schmidt law:
\begin{equation}
M_{\star,\mathrm{new}} = \epsilon \frac{M_\mathrm{gas}}{t_\mathrm{ff}} \Delta t \, ,
\end{equation}
where $t_\mathrm{ff}=\sqrt{3\pi/32\G \rho_\mathrm{gas}}$ is the free-fall time, $\rho_\mathrm{gas}~=~M_\mathrm{gas}/\Delta x^3$ is the gas density in the cell, and $\epsilon$ depends on the local turbulence of the gas, as detailed in \cite{Trebitsch_18}. 

For the feedback of supernovae, we use the mechanical feedback described in \cite{Kimm_14}, in which star particles older than 5 Myr release $\eta_\mathrm{SN}\times 10^{50}\erg/\Msun$,  where $\eta_\mathrm{SN}=0.2$. The amount of energy and momentum deposited depends on local properties of the gas (density and metallicity) so that it captures either the Sedov or the supernovaeow-plough expansion phase of the explosion.

We use the model of BHs described in \cite{Dubois_12}, where accretion is computed using the Bondi-Hoyle-Littleton formalism capped at the Eddington luminosity. AGN feedback consists of a dual-mode approach, where thermal energy, corresponding to 15\% of the bolometric luminosity (with radiative efficiency of $\epsilon_{\rm r}=0.1$), is injected at high accretion rates (luminosity above 0.01 the Eddington luminosity); otherwise feedback is modeled with a bipolar jet with a velocity of $10^4\,\rm km\, s^{-1}$ and an efficiency of 100\%.  
We slightly modify the implementation of BH dynamics: in the original \texttt{Ramses} version, the mass of the BH is deposited onto the so-called ``cloud'' particles, which uniformly pave a sphere of $4\Delta x$ radius on a grid of $\Delta x/2$ inter-cloud distance. This has the effect of smoothing the density, and therefore, when two BHs pass close by, their potential is shallower than it should be, and this delays the formation of the binary. We simply deposit all the mass of BHs onto their central cloud particle and then perform the cloud-in-cell to obtain more accurate dynamics, while using the rest of cloud particles to compute the Bondi-Hoyle-Littleton accretion rate onto the BH.

We include the dynamical friction implementation as described in Section~\ref{Subgrid_model_for_dynamical_friction}
when necessary, and dynamical friction from gas was already included \citep{Dubois_14} using Eq.~(12) from \cite{Ostriker_99}. Two BHs are allowed to merge when they are separated by less than $4\Delta x$ and the kinetic energy of the binary is lower than the gravitational energy.

\hugo{Due to our inability to resolve the cold and dense regions of the interstellar medium, gas dynamical friction and gas accretion, which depend linearly on the gas density, $\rho_{\rm gas}$ can be underestimated in simulations. To correct for this lack of resolution,  \cite{Booth_09, Dubois_14} compute the expressions obtained analytically (Chandrasekhar gas dynamical friction and Bondi accretion) and boost them by:
\begin{eqnarray}
\rm {boost}=\left(\frac{\rho_{\rm gas}}{\rho_{\rm th}}\right)^\xi \, ,
\label{eq:boost}
\end{eqnarray}
where $\rho_{\rm th}$ is a free parameter, similar to that used for star formation, which is linked to the Jeans length and depends on resolution. This parameter is typically calibrated via the phase diagram of gas, and it is $\sim 1\,\amu \cm^{-3}$ for $\Delta x=100$~pc and $\sim 50\,\amu \cm^{-3}$ for $\Delta x=10$~pc. $\xi$ can differ for accretion and dynamical friction, for the rest of the paper, we will use $\xi=\alpha$ when we relate to boosting accretion, and $\xi=\beta$ when we refer to gas dynamical friction. \cite{Booth_09} performed a parameter study and found that $\alpha = 2$ is the optimal value to recover the BH mass - galaxy mass relation \citep{Kormendy_13}. Although the expression of these boosts is not physically motivated, they have been proven to give excellent match with observations, as shown by large cosmological simulations \citep{Dubois_14b}. For this work we will either use $\xi = 2$ either $\xi = 0$ (no boost), and in Section~\ref{sec_IdealGalaxy} we also vary $\rho_{\rm th}$ within an order of magnitude of the typical value expected for the chosen resolution. We briefly explore the effects of the boost in \S \ref{sec_IdealGalaxy}.}

\section{Validation of the dynamical friction implementation}
\label{sec:ValidationOfThedynamical frictionImplementation}

\subsection{Isolated dark matter halo}
\label{subsec:Isolateddark matterHalo}
In order to compare the dynamical friction timescale with analytical estimates \citep{Lacey_93, Colpi_99, Taffoni_03}, we test our implementation following the dynamics of a BH moving in a dark matter halo.

The dark matter halo, initialized with \texttt{DICE} \citep{Perret_16}, follows a Navarro, Frank and White \citep[NFW,][]{1997ApJ...490..493N} profile with a total virial mass $M_\text{vir}=2\times 10^{11}\Msun$, a concentration parameter of $4$ and a virial radius $R_\text{vir}=45\kpc$, typical of redshift 3. We set the total spin parameter to $0.04$ consistent with the average spin parameter of cosmological dark matter halos \citep{Bullock_01}\hugo{, which only mildly evolves between $z=3$ and today \citep{MunosCuartas_14, Ahn_14}}. The BH mass is set to $10^8 \Msun$, it is initially $5\kpc$ away from the center with a tangential velocity of $57 \km \s^{-1}$, corresponding to to 50\% of the circular velocity. In the simulation, the influence radius varies between 10 and 100 pc: it is  at best resolved by 2 cell elements, therefore dynamics is generally not treated properly and dynamical friction must be added \emph{ad hoc} with our subgrid model when necessary.

As done in \cite{Tremmel_15}, we  can estimate the goodness of the method by comparing it to the analytical estimate of the ``sinking time'', $\tau_{\rm DF}$, defined as the time it will take for a \emph{satellite} to sink to a \emph{target}, using  Eq.~(12) from \cite{Taffoni_03}:
\begin{eqnarray}
\tau_{\rm DF} & =& 0.6\frac{r_c^2 v_c}{\G M_s} \log^{-1}\left(1+\frac{M_\textrm{vir}}{M_s} \right)\left( \frac{J}{J_c}\right)^\alpha  \\
& \simeq & 1.4 \Gyr \left( \frac{r_c}{100\pc}\right)^2 \left( \frac{v_c}{10 \km \s^{-1}}\right) \left( \frac{10^4 \Msun}{M_s}\right)  \nonumber \\ 
& & \log^{-1}\left(1+\frac{M_\textrm{vir}}{M_s} \right)\left( \frac{J}{J_c}\right)^\alpha \, ,\label{eq:taudynamical friction}
\end{eqnarray}
where $M_\textrm{vir}$ is the virial mass of the \emph{target}, $v_c$ is the circular velocity at the virial radius, $\G$ the gravitational constant, $r_c$ is the radius at which a test particle moving in the potential of the \emph{target} has the same energy as the \emph{satellite}, $M_s$ is the mass of the \emph{satellite}, $J$ is the specific angular momentum of the \emph{satellite} in the frame of the \emph{target}, $J_c$ is the specific angular momentum of circular orbit at $r_c$ and $\alpha$ depends on $M_s$, $M_\textrm{vir}$, $R_\textrm{vir}$ and $r_c$ and is given by Eq.~(15) from \cite{Taffoni_03}. In this case, the target is the halo and the satellite is the BH. Using this approach, we find that the BH should sink in the potential well of the halo in $600 \Myr$.

We perform two simulations which only differ by the presence (\texttt{PD}), or not (\texttt{NoDrag}), of dynamical friction onto the BH using our subgrid model. In both cases, the size of the box is  $100\kpc$, slightly larger than $2 R_\text{vir}$ and we allow refinement from levels $7$ to $11$, leading to a maximum physical resolution of $\Delta x = \Delta x_{\rm DM} = 50 \pc$, similar to what simulations reach in cosmological zooms \citep{Dubois_14}. The refinement is done using a quasi-Lagrangian criterion: a cell is refined if its mass exceeds $8\times m_\text{DM}$, where $m_\text{DM}$ is the mass of dark matter particles, and we refine at maximum level up to $4\Delta x$ around the BH. We set the mass of dark matter particles to $10^5 \Msun$, in good agreement with the value suggested by \cite{Power_03}: $m_\text{DM}=M_\text{vir}\left( R_\text{vir}/\Delta x \right)^{-2}\sim 2\times  10^5 \Msun$.

\begin{figure}
\includegraphics[width=\columnwidth]{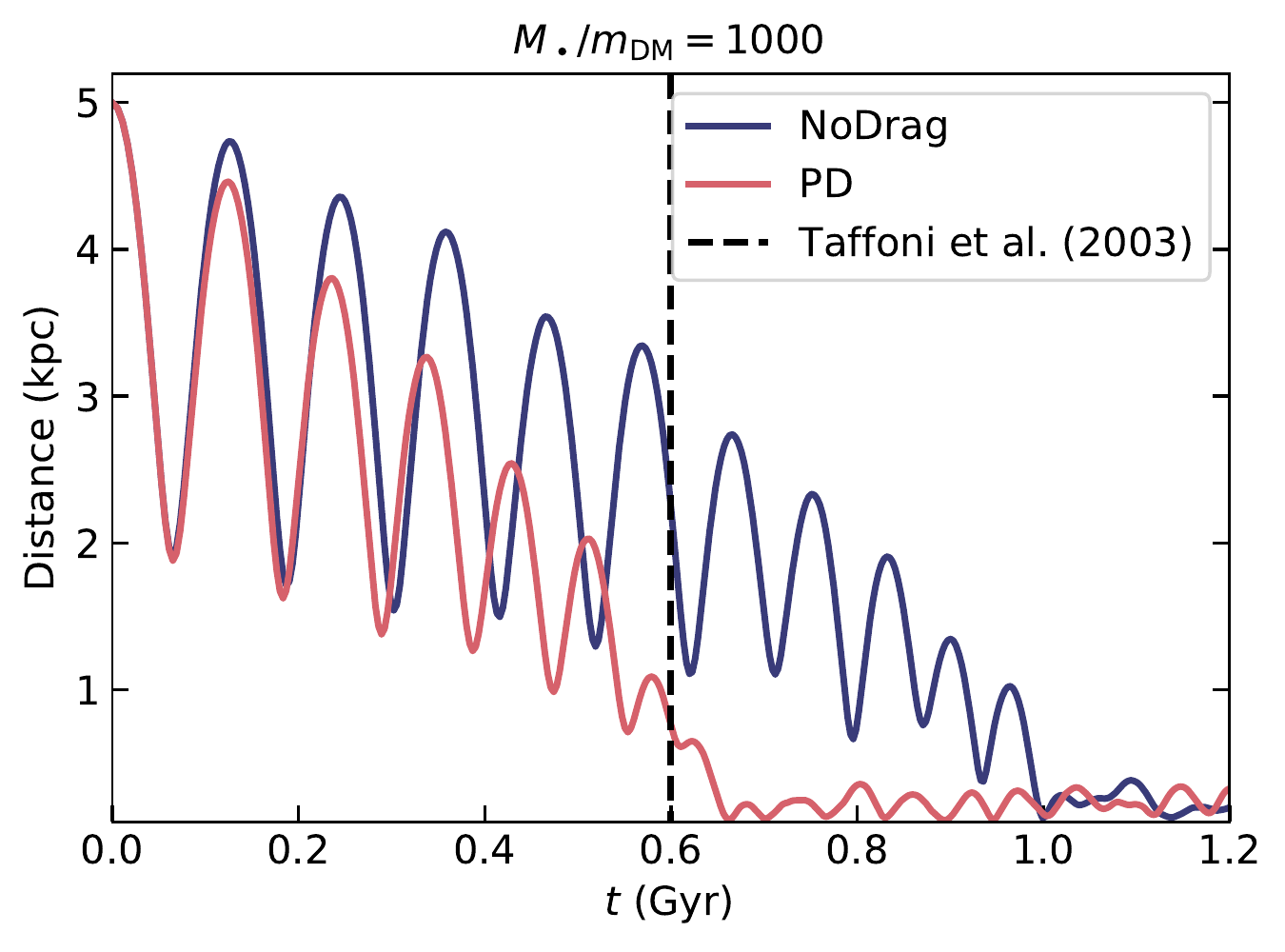}
\caption{Distance of the BH to the center of the halo as a function of time. The vertical dashed line is the analytical estimate of the total sinking time from \protect\cite{Taffoni_03}. With our prescription, the sinking time is much shorter than without and in very good agreement with the analytical estimate. The resolution in these simulations is 50 pc and the BH to dark matter mass particle ratio set equal to 1000. See section~\ref{subsec:Isolateddark matterHalo} for details.}
\label{fig_DofT_halo}
\end{figure}

We show in Fig.~\ref{fig_DofT_halo}, for both simulations, the distance between the BH and the center of the halo; we also include for comparison the analytical estimate given by Eq.~\eqref{eq:taudynamical friction}. The result is quite clear and in agreement with \cite{Tremmel_15}: adding unresolved dynamical friction contributes to recover sinking times estimated analytically. In the following section, we set in a more realistic problem where a BH sinks in a galaxy including not only dark matter but also gas, stars and many associated processes (cooling, star formation, supernovae feedback).

\subsection{Isolated galaxy}
\label{sec_IdealGalaxy}

\begin{figure}
\includegraphics[width=\columnwidth]{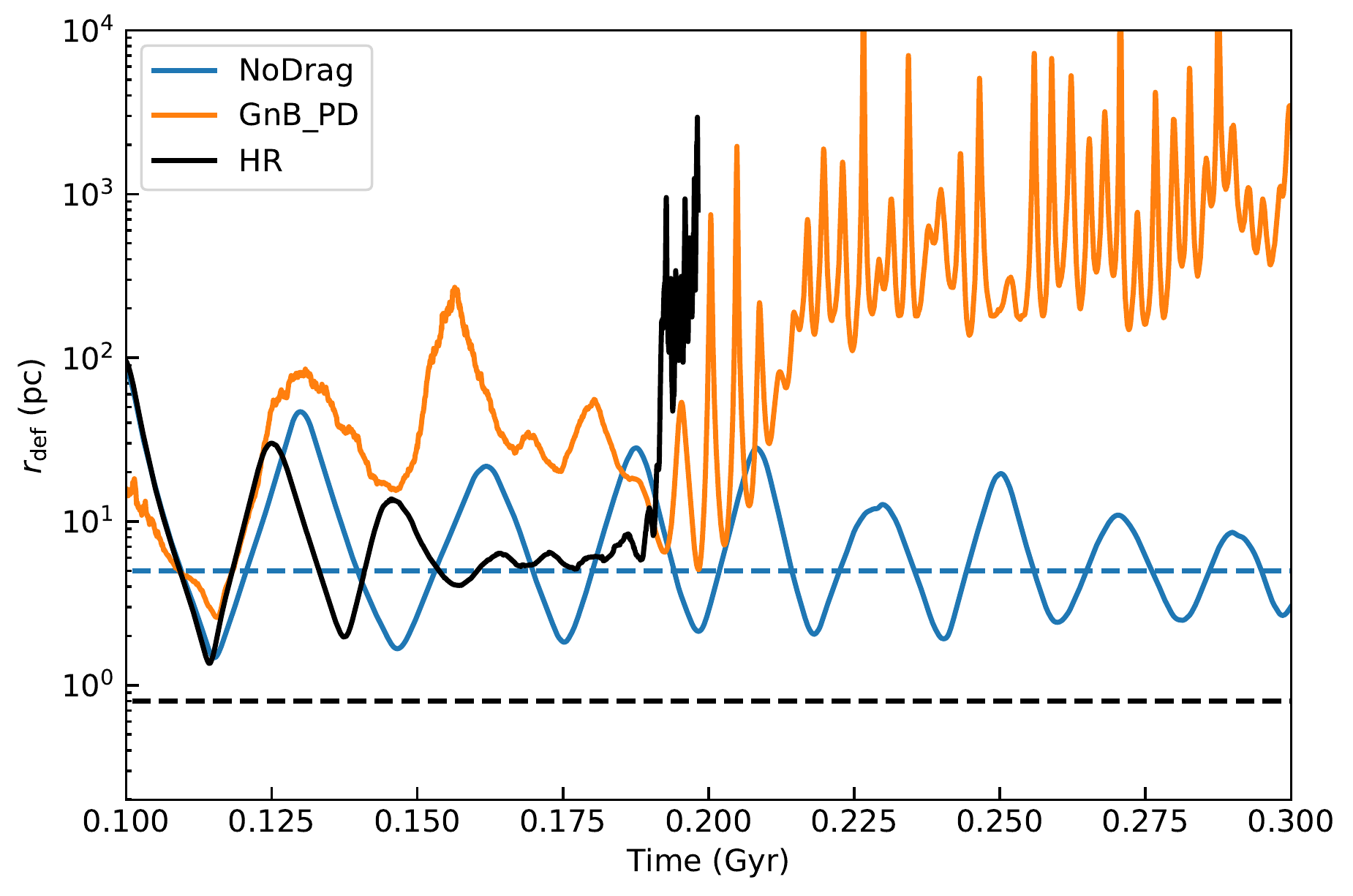}
\caption{Deflection radius (solid line) and resolution of the different simulations (dashed lines) of idealized isolated galaxies. In the low-resolution case the deflection radius is not always resolved, leading to incorrect dynamics of the BH and the need to add unresolved dynamical friction. In the high resolution run the deflection radius is always resolved and dynamical friction is self-consistently captured by the gravity solver. All quantities shown as a function of time.}
\label{fig_InfRad_gal}
\end{figure}

We run a suite of simulations (see Table \ref{Table_sims_gal}) of a BH sinking in the potential well of an idealized isolated galaxy. Our suite contains low-resolution ($\Delta x = \Delta x _{\rm DM} = 50\pc$) simulations, similar to what high-resolution zoomed cosmological simulations can reach today. Thus it is a good test to see how our implementation will act in this context. Contrary to the dark matter halo case, we do not have analytical estimates to provide a benchmark. To overcome this issue, we run a high resolution test ($\Delta x = \Delta x _{\rm DM} = 1\pc$) to perform the comparison. The setup is chosen such that, with 50~pc resolution, during the sinking, the deflection radius, as defined in Eq.~\eqref{eq:InfluenceRadius}, is not always resolved (see Fig. \ref{fig_InfRad_gal}). In this case, dynamics is not properly treated and dynamical friction must be added \emph{ad hoc} with our subgrid model. Conversely, with 1~pc resolution the deflection radius is always resolved during the sinking and dynamical friction is well captured by the gravity solver of \texttt{Ramses}, thus providing the correct dynamics.

\begin{table}
\begin{center}
 \begin{tabular}{lccccc}
  \hline
	Name&
	Dynamical friction&
	$\rho_\text{th} ^{[1]}$&
	$\beta^{[2]}$&
   $\Delta x^{[3]}$\\
	
	&
	&
	amu cc$^{-1}$&
	&
	pc
	\\
	
	\hline
	\hline
	
	\texttt{NoDrag}&
	none&
	X&X&
	50
	\\	
  	
	\texttt{GnB\_nPD}&
	gas (no boost)&
	X&0&
	50
	\\	
	  
	\texttt{GB0.1\_nPD}&
	gas (boost)&
	0.1&2&
	50
	\\			  
	  	
	\texttt{GB1\_nPD}&
	gas (boost)&
	1&2&
	50
	\\		  	
	  	
	\texttt{GB15\_nPD}&
	gas (boost)&
	15&2&
	50
	\\	
	
	\texttt{GnB\_PD}&
	gas (no boost)+stars+dark matter&
	X&0&
	50
	\\

	\texttt{HR}&
	none&
	X&X&
	0.76
	\\		  	
	  	
  	\hline
 \end{tabular}
 \end{center}
 \caption{Different simulations performed for the isolated galaxy test, with their name and the use, or not, of our new model. [1-2] Typical density and exponent that we used to boost gas friction, following eq. \eqref{eq:boost}. [3] Resolution of the simulation.}
 \label{Table_sims_gal}
\end{table}

We initialize with \texttt{DICE} an ideal galaxy at redshift 3 with a total virial mass of $2\times 10^{11}\Msun$ and a spin parameter of $0.04$. The galaxy is composed of four components.
\begin{itemize}
\item A dark matter halo with a mass of $1.95\times 10^{11} \Msun$, slightly lighter than in \S \ref{subsec:Isolateddark matterHalo}. It has a virial radius of $45\kpc$ and the density follows a NFW profile with a concentration parameter of $4$.
\item A gas disk with a total mass of $2.4\times 10^9 \Msun$. The density follows an exponential disk + sech-$z$ profile with a scale radius of $1.28\kpc$ and an aspect ratio of 1:10. We impose an initial constant absolute metallicity and temperature of $10^{-3}$ and $10^5\K$, respectively.
\item A stellar disk with a total mass of $1.6\times 10^9 \Msun$. The density follows an exponential disk + sech-$z$ profile with a scale radius of $1.28\kpc$ and an aspect ratio of 1:10. We impose an initial constant absolute metallicity of $10^{-3}$. Additionally, to avoid unphysical initial starbursts regularly found in ideal simulations \citep{Capelo_15}, we give an age distribution to stellar particles to mimic a 5 $\Msun \yr ^{-1}$ star formation rate.
\item A stellar bulge with a total mass of $8\times 10^8 \Msun$. The density follows a Hernquist profile \citep{Hernquist_90} with a scale radius of $0.128\kpc$. We impose a constant absolute metallicity of $2\times 10^{-4}$ (5 times smaller than in the disk to mimic the older age of stars in the bulge). Similarly, we give an age to stellar particles to mimic a 0.5 $\Msun \yr ^{-1}$ star formation rate. 
\end{itemize}

In the low-resolution simulations (50 pc), the mass of dark matter particles is set to $10^6~\Msun$ and that of star particles to $2\times 10^4 \Msun$. In the high resolution simulations (1 pc) the mass of dark matter particles is set to $5\times10^4 \Msun$ and that of star particles to $2\times 10^3 \Msun$. In both cases the size of the box is 100 kpc and we allow for refinement from levels 7 to 11 in the low resolution simulations and from 7 to 17 in the high resolution one, refining the mesh when  $M_\mathrm{DM}^\mathrm{cell}+ 10 M_b^\mathrm{cell} \geq 8 m_\mathrm{DM} $, where $M_\mathrm{DM}^{\rm cell}$ and $M_b^\mathrm{cell} $ are, respectively, the mass of dark matter and baryons in the cell. Maximum refinement is enforced within $4\Delta x$ around the BH.

After initializing this galaxy, we switch on cooling, star formation, supernovae feedback (see \S \ref{section:SubgridPhysics}) and let the galaxy relax for 100 Myr. At that point, a BH with mass $10^7 \Msun$ is placed in the $z=0$ plane, 1 kpc from the center and with a tangential velocity of 21 km/s, corresponding to 30\% of the circular velocity. Accretion and feedback from the BH are not included in order to keep the BH mass constant and isolate the effects of dynamical friction. We include dynamical friction with different implementations: from collisionless particles and gas without boost (\ie no free parameters), or only from gas, with or without a boost factor. The simulation properties and set-up are summarized in Table \ref{Table_sims_gal}.

\begin{figure}
\includegraphics[width=\columnwidth]{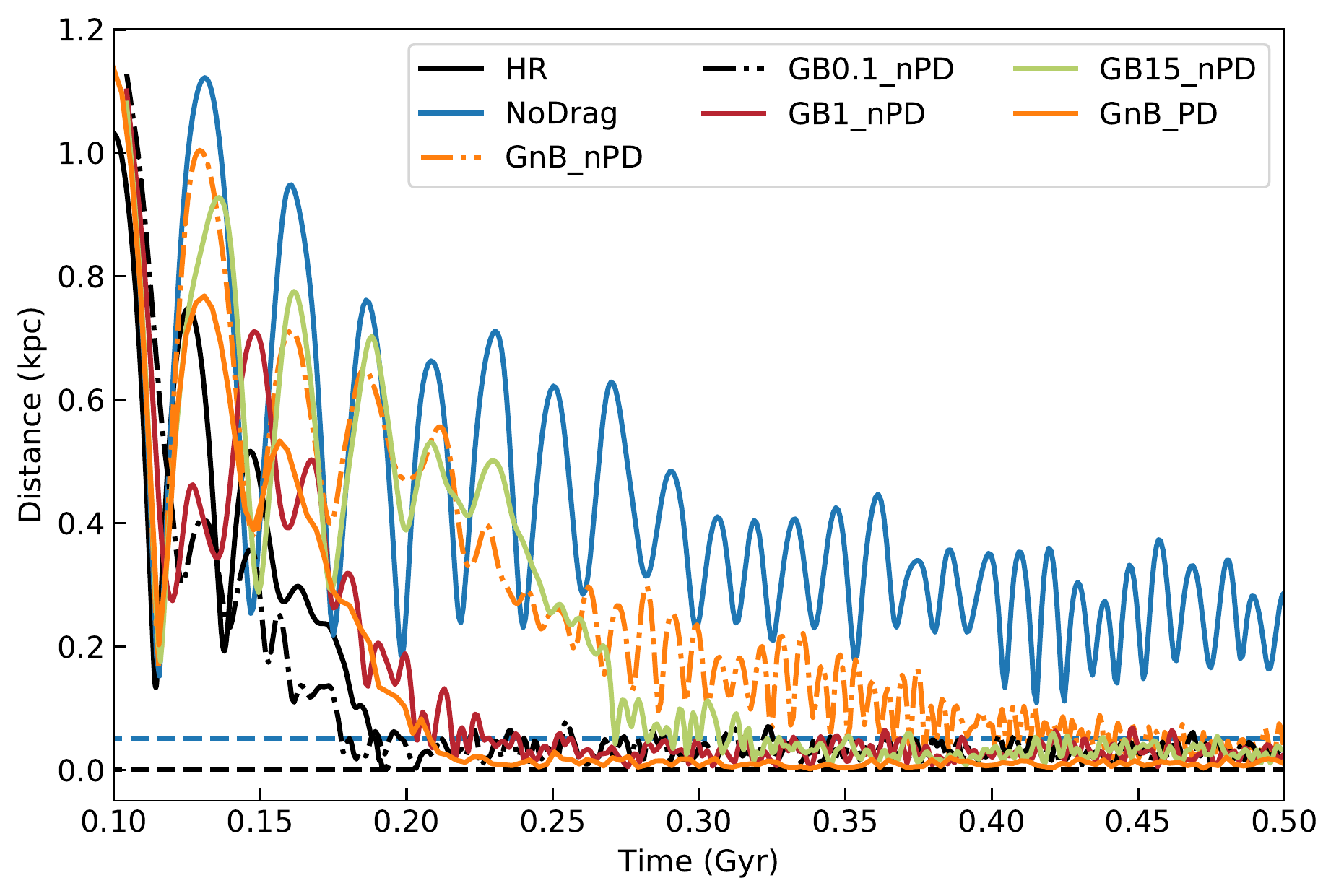}
\caption{Distance of the BH to the center of the galaxy as a function of time and resolutions of the different simulations (dashed lines). The good behavior of our model is confirmed by the agreement between the solid orange curve (low resolution, use of our model) and the solid black one (high resolution). See section~\ref{sec_IdealGalaxy} for details.} 
\label{fig_DofT_gal}
\end{figure}

We show in Fig. \ref{fig_DofT_gal}, for all our simulations, the distance between the BH and the center of the galaxy as a function of time. We first stress the difference between low resolution simulations with gravity only, \ie without including the dynamical friction model (\texttt{NoDrag}, blue line), and the simulations at high resolution where the deflection radius is resolved (\texttt{HR}, black line). In agreement with the results of \cite{Pfister_17}, resolving at least the deflection radius is mandatory to properly capture the dynamics of the BH in the dynamical friction phase.

We now compare simulations where we vary $\rho_\mathrm{th}$ (\texttt{GB0.1\_nPD}, \texttt{GB1\_nPD}, \texttt{GB15\_nPD}) but we do not include dynamical friction from stars and dark matter. As expected, the lower $\rho_\mathrm{th}$, the larger the boost, the faster the BH sinks. The choice of $\rho_\mathrm{th}$ must be performed accurately: if $\rho_\mathrm{th}$ is too low, BHs can get caught in a passing clump and either follow the clump outside the galaxy center, or remain artificially  in a dense environment where accretion is triggered, resulting in an overestimate of the mass of BHs.  If $\rho_\mathrm{th}$ is too high, instead, the correction to dynamical friction is insufficient and the orbital decay is delayed. In this particular case, $\rho_\mathrm{th}$ between 0.1 and 1$ \amu \cm^{-3}$ is the best value to recover the high resolution results, but the exact value may depend on additional factors such as the gas fraction (50\% in our case). 

We finish with the simulation where the influence radius is not always resolved, but in which we include sub-grid dynamical friction from stars, dark matter and gas (without any boost) following our implementation (\texttt{GnB\_PD}, orange line). This implementation does not contain any free parameters and avoids the arbitrary choice of $\rho_\mathrm{th}$. This simulation is in excellent agreement with the high resolution simulation  (\texttt{HR}, thin black line), confirming the good behavior of our model in a realistic, although idealized, galaxy. 

\subsection{Limits of the model: low mass black holes}
\label{sec_LimitsOfTheModelLowMassBHs}

In this section we explore the limits of our implementation when a BH has a mass so low that 2-body interactions with star and dark matter particles significantly perturb its dynamics.

We run simulations similar to those described in \S \ref{subsec:Isolateddark matterHalo} but decreasing the mass of the BH down to the mass of dark matter particles ($m_{\rm DM}=10^5\Msun$). To contain computational costs, we also change the orbital parameters of the BH such as the analytical estimates from \cite{Taffoni_03}, $\tau_\mathrm{DF}$, remains a few Gyrs. We list the parameters of the simulations in Table~\ref{table:LowMassBH}.

\begin{figure}
\includegraphics[width=\columnwidth]{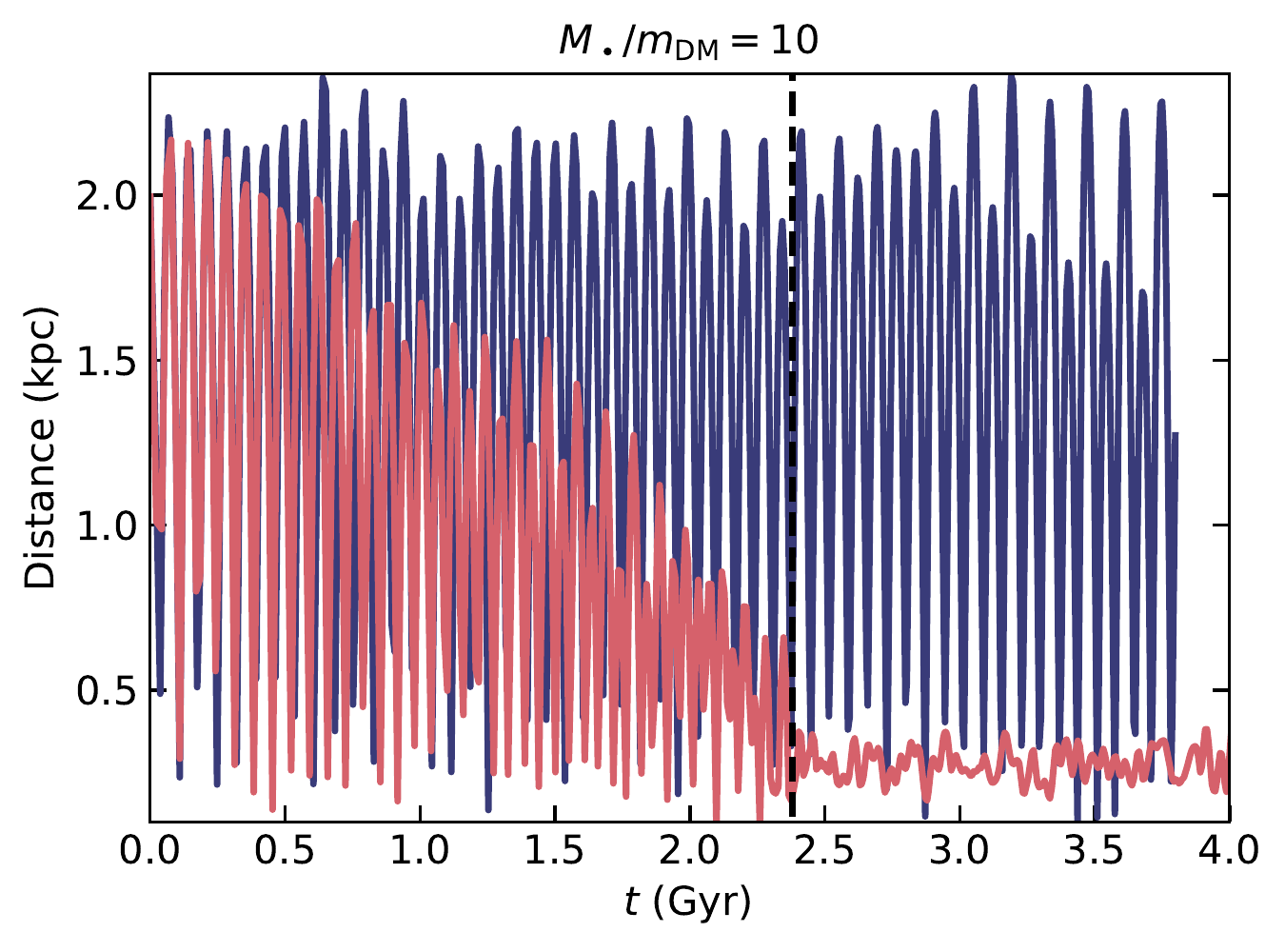}\\
\includegraphics[width=\columnwidth]{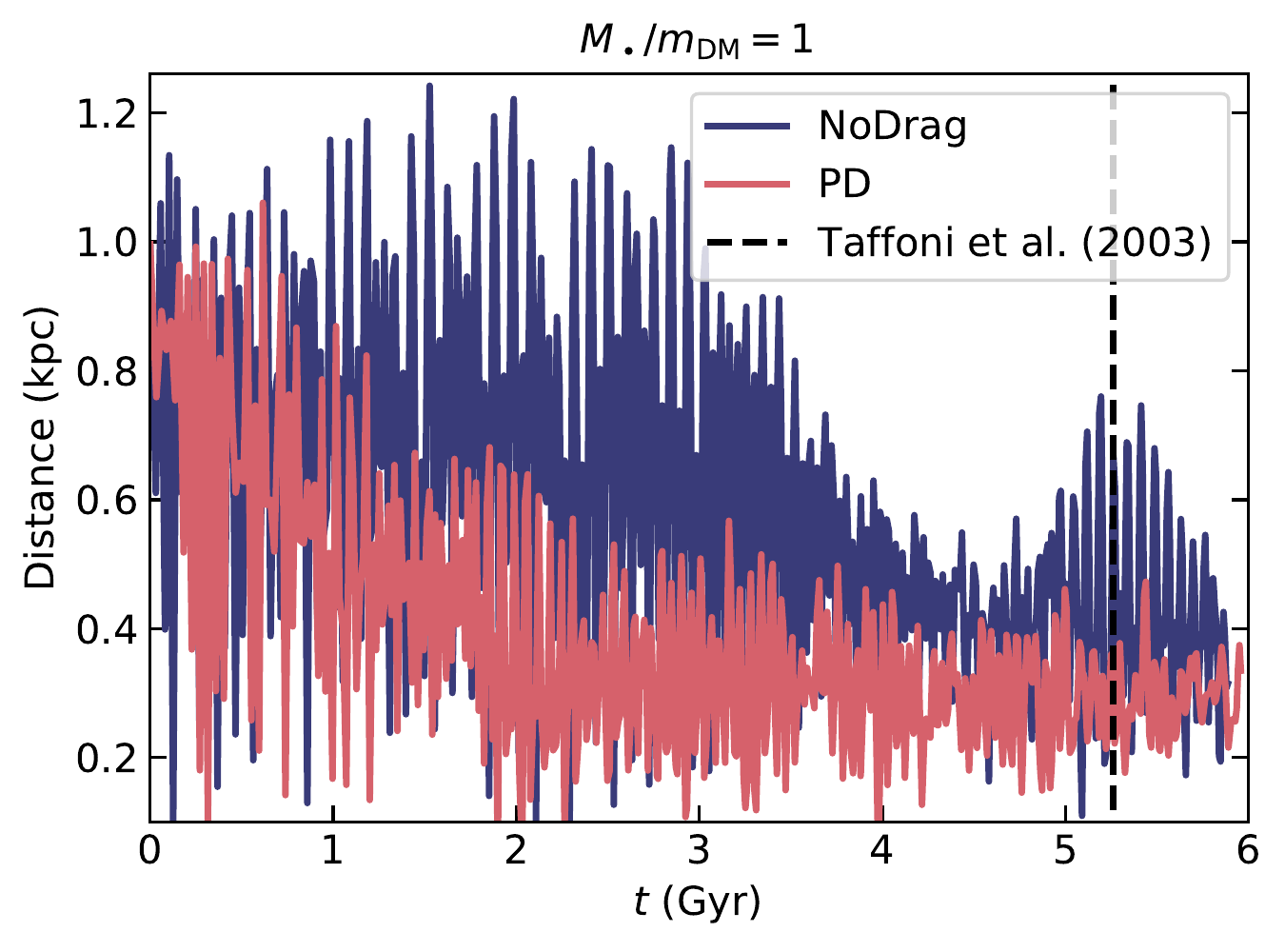}\\
\caption{Different simulations performed to test the effect of reducing the mass ratio between the BH and dark matter particles. We indicate the use (\texttt{PD}) or not (\texttt{NoDrag}) of our prescription for dynamical friction. If the BH mass is similar to the dark matter particle mass, the efficacy of the model becomes limited. See section~\ref{sec_LimitsOfTheModelLowMassBHs} for details.}
\label{fig_DistLowMBH}
\end{figure}

We show in Fig.~\ref{fig_DistLowMBH} the distance of the BH to the center of the halo as a function of time. It is clear that our model works very well when BHs have a mass larger than 10 times the mass of particles causing dynamical friction. If the mass of the BH is similar to that of particles causing dynamical friction, however,  it is scattered through 2-body interactions and the model becomes less reliable, as also noted by \cite{Tremmel_15}.

In \S \ref{sec_IdealGalaxy} and \S \ref{section:CosmologicalSimulations}, the mass of dark matter particles is larger than that of BHs. However, we use cloud-in-cell interpolation to smooth the dark matter distribution, and we ensure that the mass of star particles, which are the main source of dynamical friction, is lower than the mass of BHs.

\begin{table}
\begin{center}
 \begin{tabular}{lccc}
  \hline
	$M_\bullet / m_\mathrm{DM}$&
	$d_0$&
    $v_0$&
    $\tau_\mathrm{DF}$\\
	
	&
	$\kpc$&
	$\km \s^{-1}$&
	$\Gyr$
	\\
	
	\hline
	\hline

	10&2&8.2&2.38
	\\
	
	1&1&6&5.26
	\\
	  	
  	\hline
 \end{tabular}
 \end{center}
 \caption{Different simulations we perform to test the limits of our model in terms of particle mass ratio. We indicate the different mass ratio between the BH and dark matter particles, the initial distance of the BH from the center of the halo, the initial velocity of the BH and the analytical estimate for the time the BH should take to reach the center of the halo from \protect \cite{Taffoni_03}. In all cases, we run a simulation with (\texttt{PD}) and without (\texttt{NoDrag}) our model.}
 \label{table:LowMassBH}
\end{table}

\section{Cosmological simulations}

\subsection{Set-up}
\label{section:CosmologicalSimulations}
\begin{table*}
 \begin{tabular}{lccccccccc}
  \hline
	Name&
	Particle&
	$\rho_\text{th} ^{[1]}$&
   	$\Delta x^{[2]}$&
   	$\Delta x^{[3]}_{\rm DM}$&
   	$m_\star ^{[4]}$&
   	$m_\text{DM} ^{[5]}$&
 	$M_\bullet ^{[6]}$\\

	&
	 dynamical friction
	&
	amu cc$^{-1}$
	&
	pc
	&
	pc
	&
	$\Msun$
	&
	$\Msun$
	&
	$\Msun$
	\\
	
	\hline
	\hline

	\\

	\texttt{LR\_PD\_BH1e4}&
	\vmark&
	5&72&2300&$2\times10^4$&$2\times 10^6$&$10^4$
	\\

	\texttt{MR\_PD\_BH1e4}&
	\vmark&
	10&36&2300&$2\times10^4$&$2\times 10^6$&$10^4$
	\\		 	
	
	\texttt{HR\_PD\_BH1e4}&
	\vmark&
	50&9&572&$2\times10^3$&$2\times 10^5$&$10^4$
	\\
	
	\texttt{HR\_nPD\_BH1e4}&
	\xmark&
	50&9&572&$2\times10^3$&$2\times 10^5$&$10^4$
	\\

	\texttt{HR\_PD\_BH1e5}&
	\vmark&
	50&9&572&$2\times10^3$&$2\times 10^5$&$10^5$
	\\
	
	\texttt{HR\_nPD\_BH1e5}&
	\xmark&
	50&9&572&$2\times10^3$&$2\times 10^5$&$10^5$
	\\		 	
	
  	\hline
 \end{tabular}
 \caption{Properties of the suite of cosmological simulations performed. [1] Typical density and exponent that we used to boost accretion, following Eq. \eqref{eq:boost}. [2-5] Spatial/mass resolution of the simulation. [6] Seed mass of BHs. }
 \label{table:CosmoSims}
\end{table*}

We now move to the full cosmological context, endeavoring to study the dynamical behavior of seed BHs in high-redshift galaxies. We run a suite of cosmological simulations, with the code \texttt{Ramses}. We zoom-in on one halo using different prescriptions for the dynamics of BHs. \hugo{As we are interested in understanding the evolution of BHs in typical galaxies, we chose a halo with a minor/major merger rate comparable to the mean evolution obtained by \cite{Fakhouri_10} in this mass range.} The physics is similar to that of the simulations described in \S \ref{section:SubgridPhysics} but for the refinement strategy: we refine if $M_\mathrm{DM}^\mathrm{cell}+ (\Omega_m / \Omega_b -1) M_b^\mathrm{cell} \geq 8 m_\mathrm{DM} $, where $M_\mathrm{DM}$ and $M_b^\mathrm{cell} $ are, respectively, the mass of dark matter and baryons in the cell, and  $\Omega_m$ and $\Omega_b$ are the total matter and baryon density. The minimum cell size, $\Delta x$ is kept roughly constant in proper physical size with redshift: an additional level of refinement is added every time the expansion factor, $a_\textrm{exp}$, decreases by a factor of two, such that the maximum level, $l_\mathrm{max}$, is reached at $a_\textrm{exp} = 0.8$. For simplicity, we further assume that $\Delta x=L_\textrm{box}/2^{l_\mathrm{max}}$, where $L_\mathrm{box}$ is the size of the box at redshift 0. Concerning the subgrid physics of BHs (see \S \ref{section:SubgridPhysics})  we use $\beta =2$ to boost accretion, gas friction is not boosted ($\alpha = 0$) and the value of $\rho_{th}$ depends on resolution. The specifications of each simulations are described in Table~\ref{table:CosmoSims}.

\subsubsection{Initial conditions}
We assume a $\Lambda$CDM cosmology with total matter density $\Omega_m=0.3089$, baryon density $\Omega_b = 0.0486$, dark energy density $\Omega_\Lambda = 0.6911$, amplitude of the matter power spectrum $\sigma_8 = 0.8159$, $n_s = 0.9667$ spectral index and Hubble constant $H_0 = 67.74 \km \s^{-1}  \Mpc^{-1}$ consistent with the Planck data \citep{Planck_15}. The initial conditions are produced with \texttt{MUSIC} \citep{MUSIC}. The box size of the simulations is $L_\mathrm{box} = 73.8 \Mpc$, with a coarse grid of $256^3$ dark matter particles corresponding to a dark matter mass resolution of $m_\mathrm{DM,coarse} = 3 \times 10^9 \Msun$. A high-resolution region is defined around a halo of $M_\mathrm{vir} = 10^{12} \Msun$ at $z = 2$ that contains only high-resolution dark matter particles (see Table 2 for the mass of high-resolution dark matter particles in each simulation) within 2 $r_\mathrm{vir}$ ($r_\mathrm{vir} = 100 \kpc$). The halo is a progenitor of a group of galaxies whose mass is $M_\mathrm{vir} = 7\times10^{12} \Msun$ at $z=0$.

\subsubsection{Finding halos and galaxies}

We construct catalogues of haloes and galaxies using the \texttt{AdaptaHOP} halo finder \citep{Aubert_04}, which uses an SPH-like kernel to compute densities at the location of each particle and partitions the ensemble of particles into sub-haloes based on saddle points in the density field. Haloes contain at least 200 dark matter particles. Galaxies are identified in the same way, and contain at least 200 stellar particles. We then construct a merger tree for halos and galaxies with \texttt{TreeMaker} \citep{Tweed_09}. 

\begin{figure*}
 \includegraphics[width=\columnwidth]{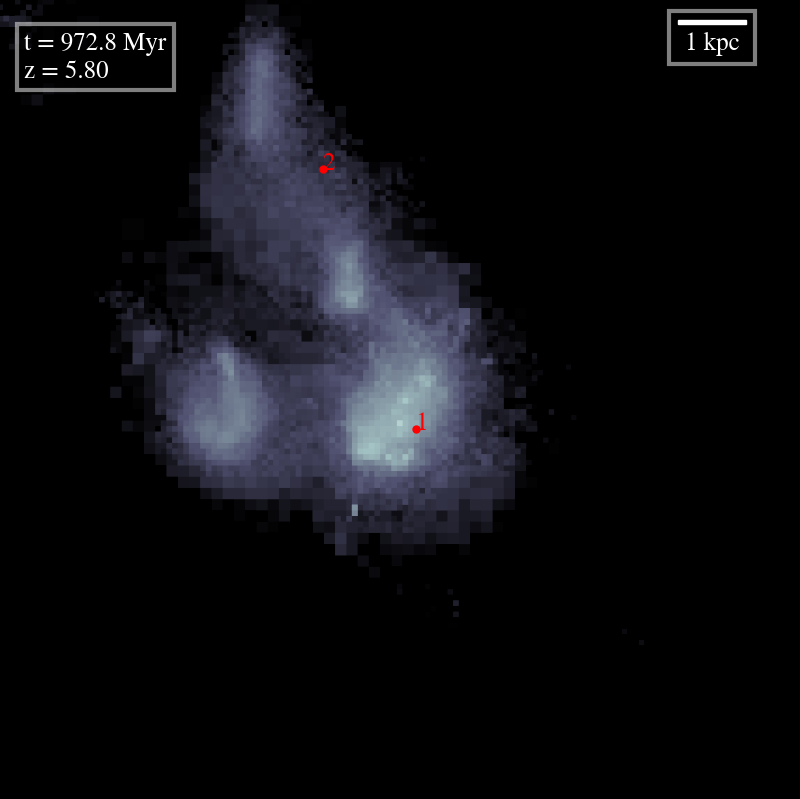}
 \includegraphics[width=\columnwidth]{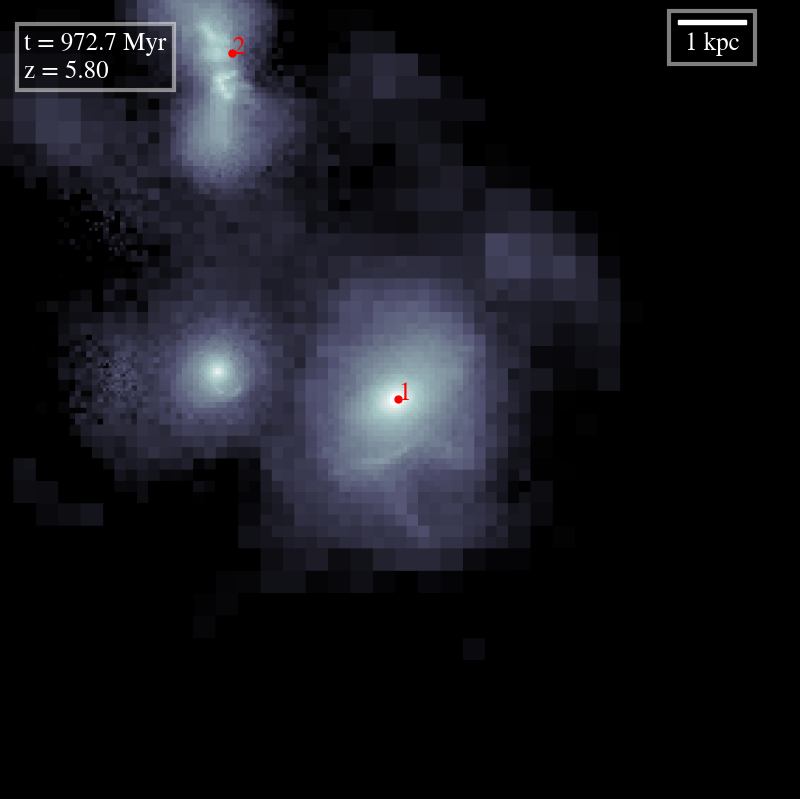}\\
 \includegraphics[width=\columnwidth]{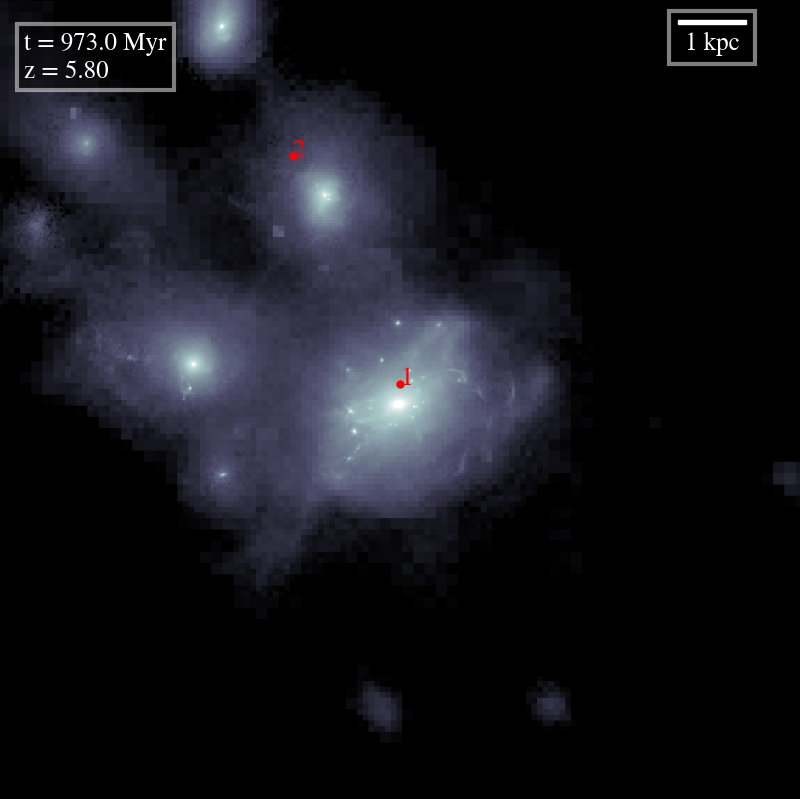}
 \includegraphics[width=\columnwidth]{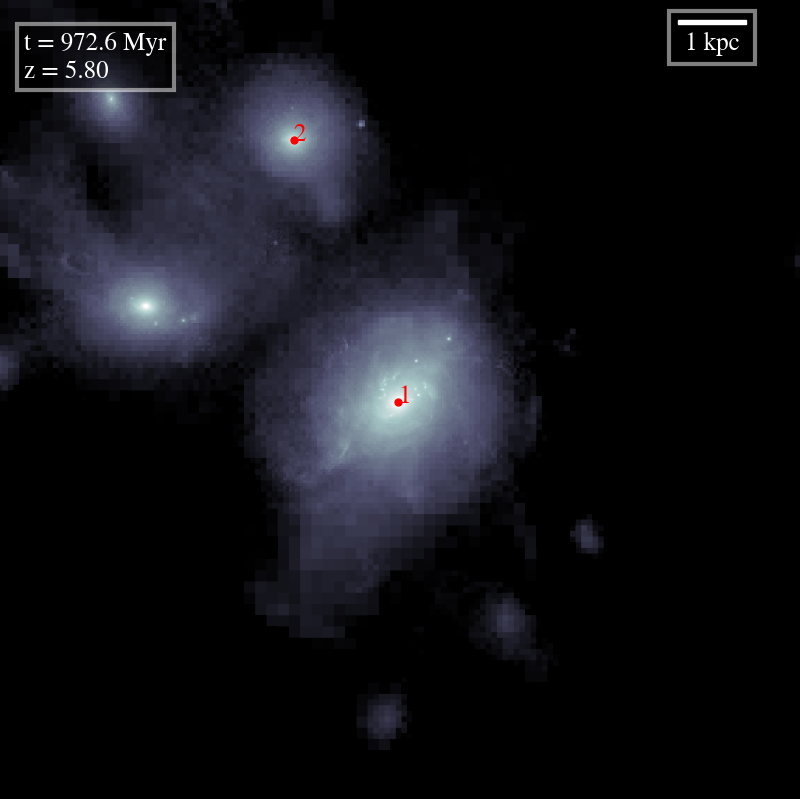}\\
 \caption{Top left: stellar density (black: $10^{-4} \Msun \pc^{-3}$, white: $1 \Msun \pc^{-3}$), centered on the main galaxy with a 80 comoving kpc box size, of \texttt{LR\_PD\_BH1e4}. Top right: \texttt{MR\_PD\_BH1e4}. Bottom left: \texttt{HR\_PD\_BH1e4}. Bottom right:  \texttt{HR\_PD\_BH1e5}. The panels show the exact same galaxy at the same time to highlight the effects of resolution. In section~\ref{section:InfallOfASeedBlackHoleInItsOwnGalaxy} we discuss the dynamics of a BH in the satellite galaxy on the top-left corner of each panel (the BH is highlighted in red and its ID is 2 in the four panels). In section~\ref{section:FormationOfABlackHoleBinaryInAHighZMerger} we discuss instead the interaction between this BH and the main BH in the central galaxy (also highlighted in red, and its ID is 1 in the four panels).}
 \label{fig:supernovaeapshots}
\end{figure*}

\subsubsection{Estimate of the sinking time}
\label{subsection:MeasurementOfTheSinkingTime}

We consider once again the ``sinking time'', $\tau_{\rm DF}$, defined as the time it takes for a \emph{satellite} to sink to a \emph{target}, using  Eq.~\ref{eq:taudynamical friction}. 
To compute $\tau_{\rm DF}$ for a BH in its own galaxy (\S \ref{section:InfallOfASeedBlackHoleInItsOwnGalaxy}), we consider that the \emph{satellite} is the BH, for which we have the dynamical properties, and we consider that the \emph{target} is the galaxy, for which we compute the different properties with the halo finder.

To compute $\tau_{\rm DF}$ for a BH during a galaxy merger (\S \ref{section:FormationOfABlackHoleBinaryInAHighZMerger}), we have to take into  account that $M_s$ evolves. Initially, the BH is surrounded by its own galaxy, which is itself surrounded by a halo, and it is the system BH+galaxy+halo that undergoes dynamical friction. Therefore, we must match BHs to galaxies and galaxies to halos to have the corrected \emph{satellite} mass, i.e. $M_s$ is similar to the mass of the halo. 

In a second phase, the dark matter halo and outer stellar layers of the secondary galaxy disperse into that of the primary, and the BH remains surrounded only by a fraction of the initial stellar mass, and we identify the evolving $M_s$ via the halo finder.  Finally, the BH remains naked, and $M_s$ is the BH mass. To give an order of magnitude for this final phase, in the early universe, where  galaxies have velocity dispersion as small as tens of $\km \s^{-1}$, unless the BH is very massive ($\gtrsim 10^5 \Msun$), or surrounded by a bound dense stellar cluster, acting as if \Ms is larger, the sinking time is longer than Gyrs if the distance to the center if larger than $\sim100 \pc$, which is likely to be the case if the BH is scattered due to anisotropies of the galaxy, either when it is in isolation or during mergers.

\subsection{Dynamics of a seed black hole in its own galaxy}
\label{section:InfallOfASeedBlackHoleInItsOwnGalaxy}

We focus on a satellite galaxy which merges with the main galaxy when the age of the Universe is about 1 Gyr. In Fig.~\ref{fig:supernovaeapshots} we show snapshots at the beginning of the interaction between the main galaxy, on which the figure is centered, and the satellite, to the top left of the main galaxy. This satellite hosts a BH and we study its dynamics while the galaxy is in relative isolation. This case is interesting because it explores the prospects for a seed BH to remain surrounded by dense cold gas available for growing the BH and make it observable as a faint AGN.

\begin{figure}
\includegraphics[width=\columnwidth]{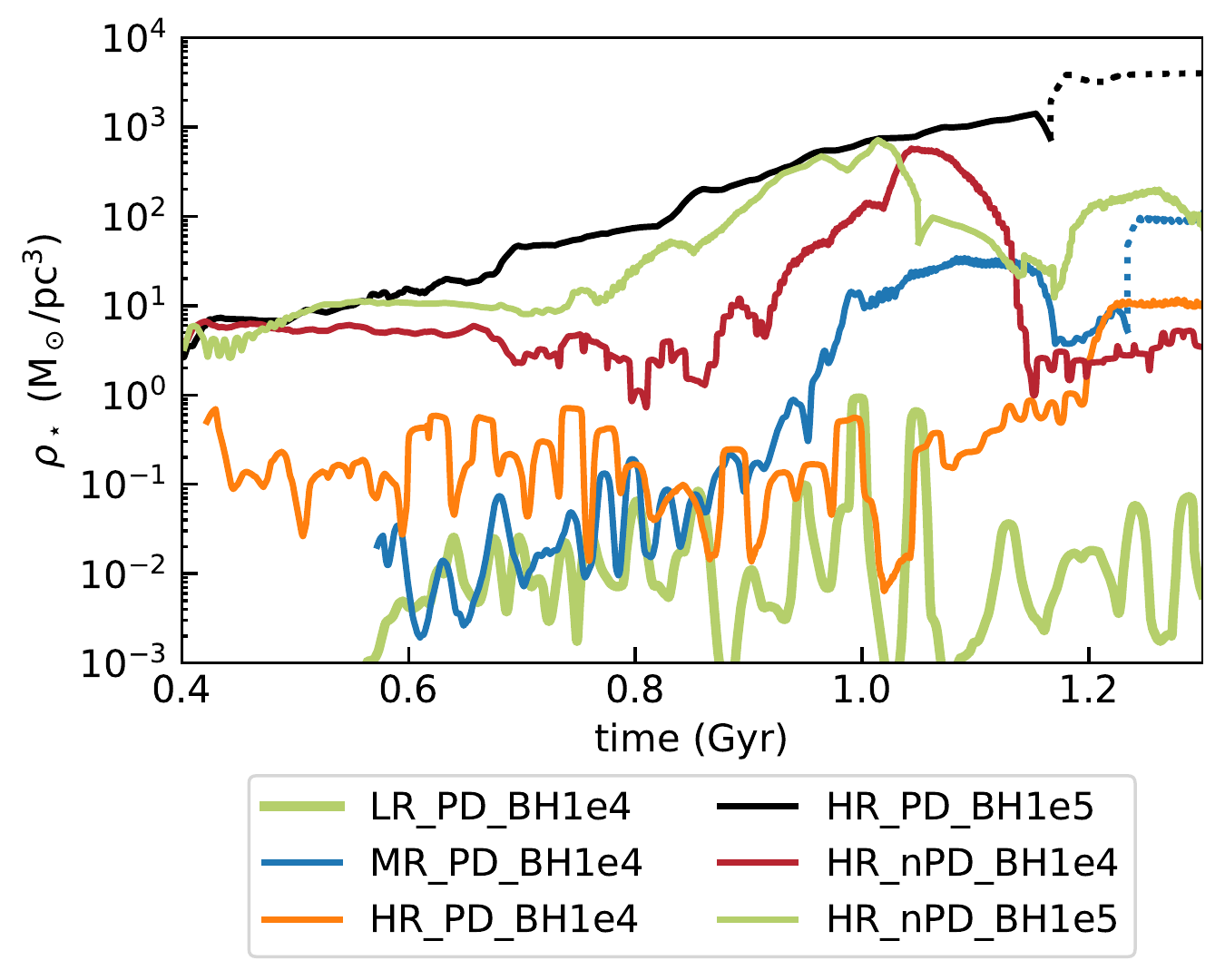}\\
\includegraphics[width=\columnwidth]{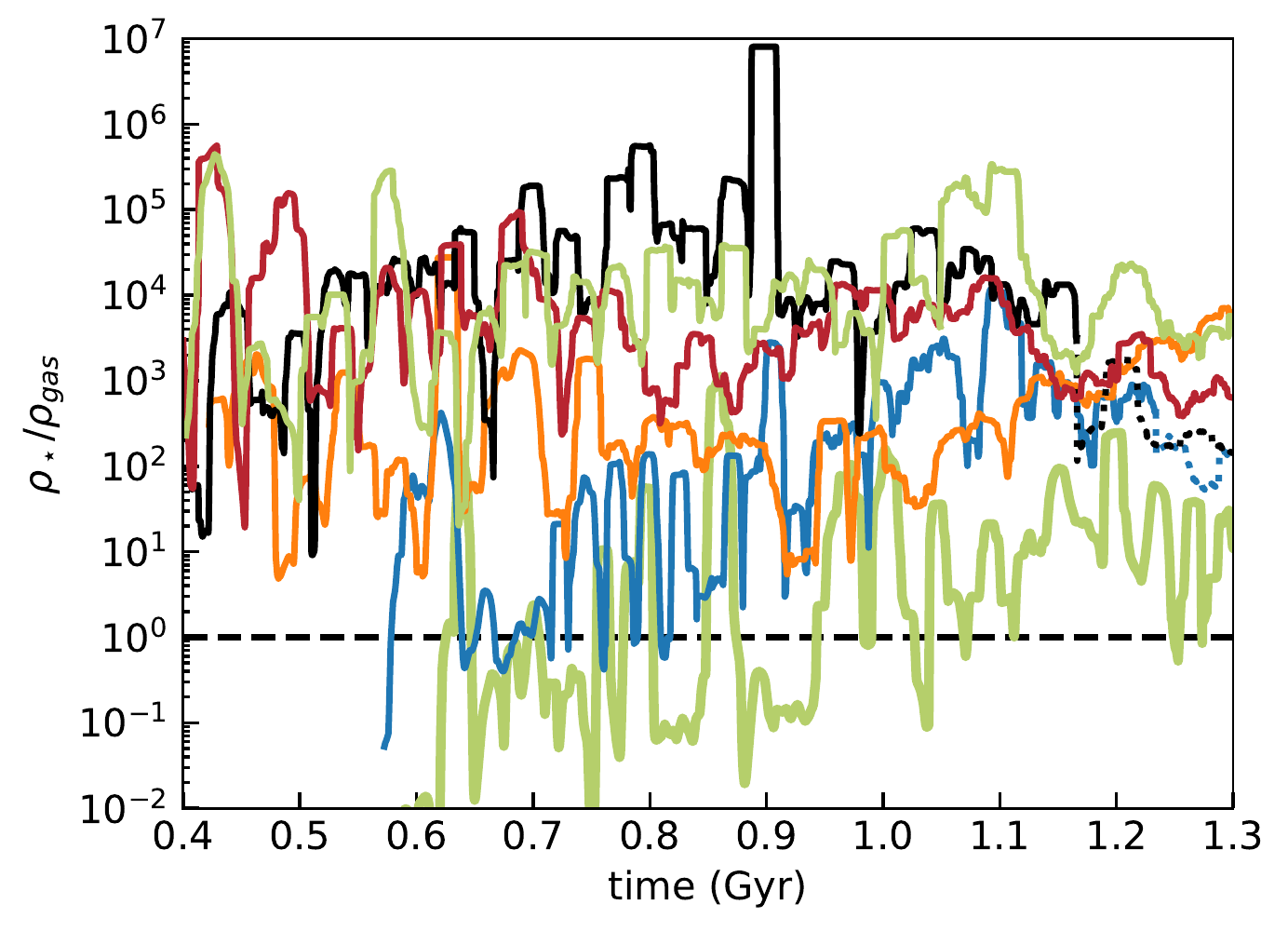}
\caption{Top panel: mean stellar density within $4\Delta x$ around the BH in the satellite galaxy, as a function of time for a subset of the simulations listed in Table~\ref{table:CosmoSims}, as noted in the inset. Bottom panel: ratio of the stellar density and gas density within  $4\Delta x$ around the satellite BH, as a function of time. }
\label{fig:RhoBH}
\end{figure}

We start by studying how the different sources of friction (dark matter, stars and gas) contribute to the dynamical evolution. Fig.~\ref{fig:RhoBH} presents the density in gas and stars around the BH (we do not include dark matter since its contribution is negligible). Gas is more chaotic than stars, but stars themselves do not provide a constant acceleration because they are also irregularly distributed. Beyond the sheer inhomogeneity, gas can shock, cool, inflow and outflow making its dynamical friction contribution unpredictable \emph{a priori}. The presence of satellites also perturbs the BH orbit when it is far from the center, see, e.g., Fig.~\ref{fig:supernovaeapshots}:  in a typical high-redshift environment a BH feels acceleration coming from different directions. 

\begin{figure}
\includegraphics[width=\columnwidth]{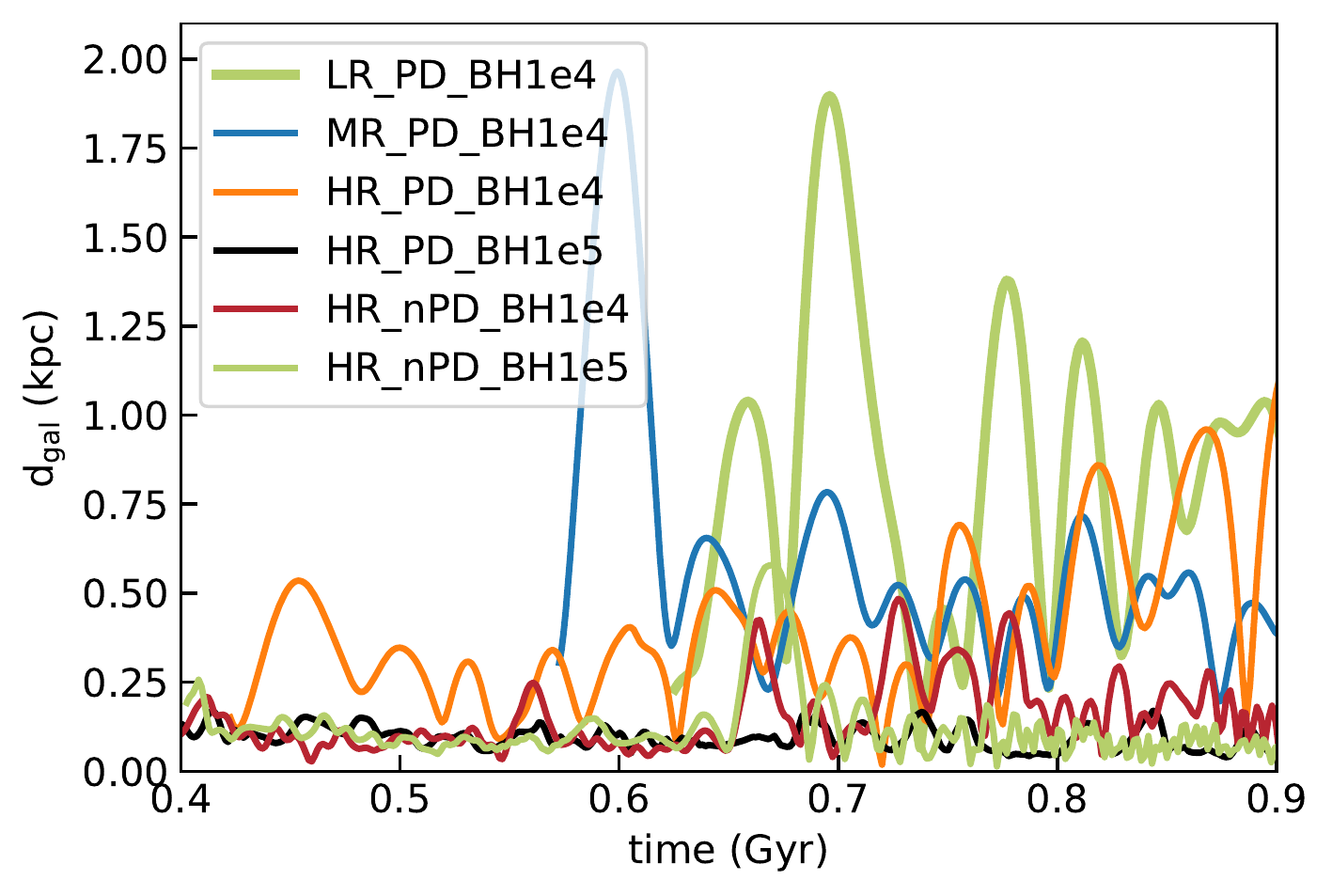}\\
\includegraphics[width=\columnwidth]{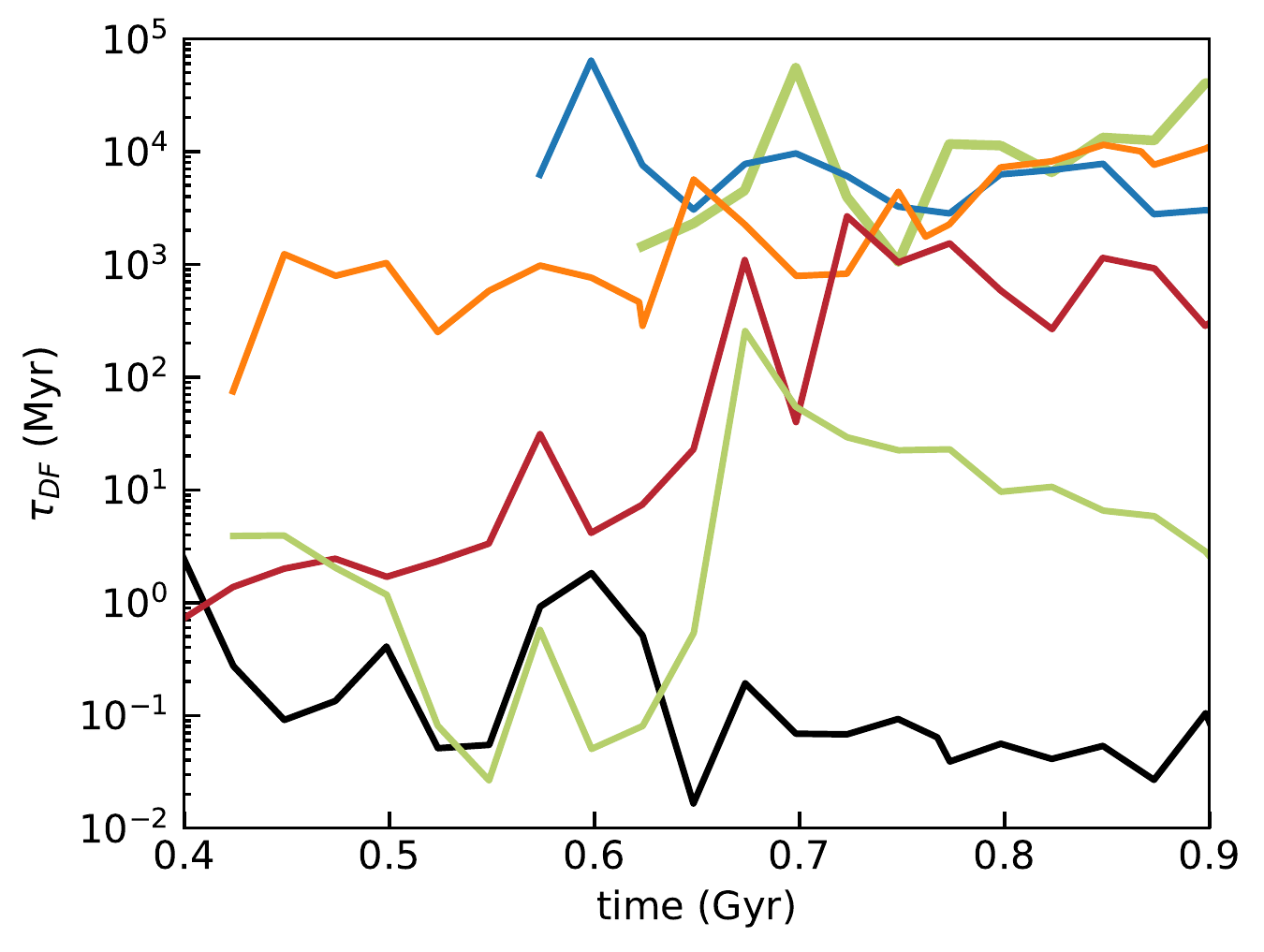}
\caption{Top panel: distance of the BH from the center of its host galaxy, before interaction with another larger galaxy, as a function of time for all simulations listed in Table~\ref{table:CosmoSims}. Bottom panel: sinking time $\tau_{\rm DF}$ for the secondary BH with respect to its host galaxy, computed using Eq.~\eqref{eq:taudynamical friction} (replacing $M_s$ by the mass of the BH in this equation), as a function of time.}
\label{fig:BHinOwnGal}
\end{figure}

Moving to how this affects the BH's orbits, we show $\tau_{\rm DF}$ as a function of time in the bottom panel of Fig.~\ref{fig:BHinOwnGal}, computed for the different simulations, using the method described in \S\ref{subsection:MeasurementOfTheSinkingTime}. We also show in the top panel of  Fig.~\ref{fig:BHinOwnGal} the distance of the BH to the center of its host galaxy.

Firstly, we find that, as long as the seeding mass of the BH is $10^4 \Msun$, all the simulations, independently of the resolution and the different models used for the BH dynamics, show a similar trend: the sinking time is, at least, 1-10 Gyr. Since in all cases $v_c$ slowly increases from 7 to 30~$\km \s^{-1}$ and $M_\bullet$ remains close to the BH seed mass, the reason of this large $\tau_{DF}$ is the dependency of the sinking time with the distance of the center of the galaxy, which is shown in the top panel of  Fig.~\ref{fig:BHinOwnGal}. Even in \texttt{HR\_PD\_BH1e4} and \texttt{HR\_nPD\_BH1e4}, where the BH is 5 times heavier than the star particles (those mostly contributing to the dynamical friction here) and forms at $\sim 70 \pc$ from the centre, it is rapidly ejected and remains hundreds of pc away from the centre.
Clumps and anisotropies are observed both in the stellar and gas central distributions. Due to such irregularities in the underlying galaxy, the BH undergoes a physically-motivated random walk out of the centre of the potential well, as it also happens in lower redshift dwarfs \citep{2019MNRAS.482.2913B}.
When the BH is more massive, $10^5 \Msun$, it remains in the center of its host, with a sinking time less than 100 Myr. $10^5 \Msun$ seems therefore to be the minimum requirement to imagine that a BH is well stabilized in the center of its host. BHs with masses lower than $10^5 \Msun$ are scattered within the galaxy due to irregularities of the gas/stellar potential and oscillate around the center of their host galaxies, remaining far from the dense gas regions, therefore we expect them to have low accretion rates \citep{Smith_18} and be difficult to observe.

\begin{figure}
\includegraphics[width=\columnwidth]{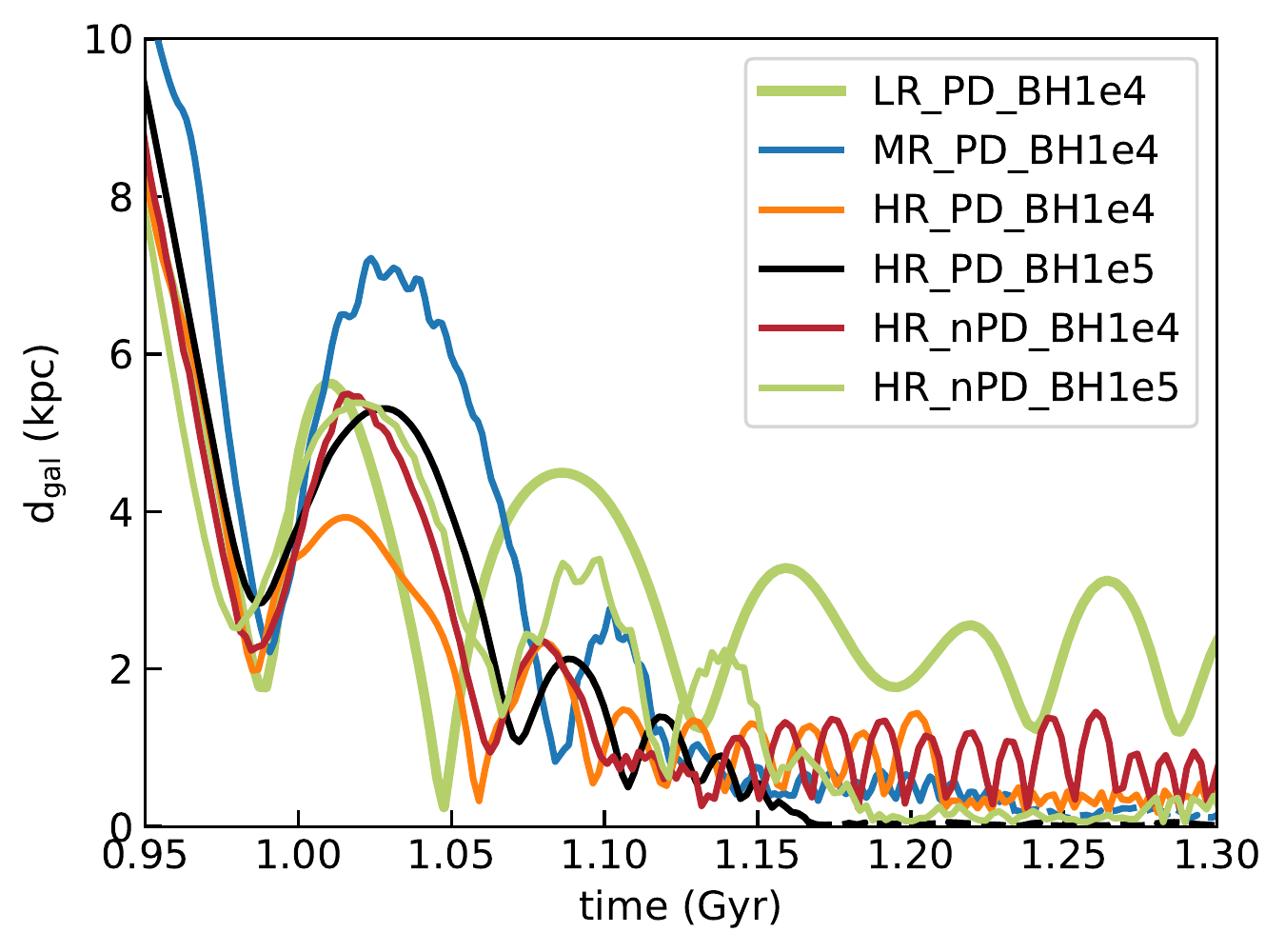}\\
\includegraphics[width=\columnwidth]{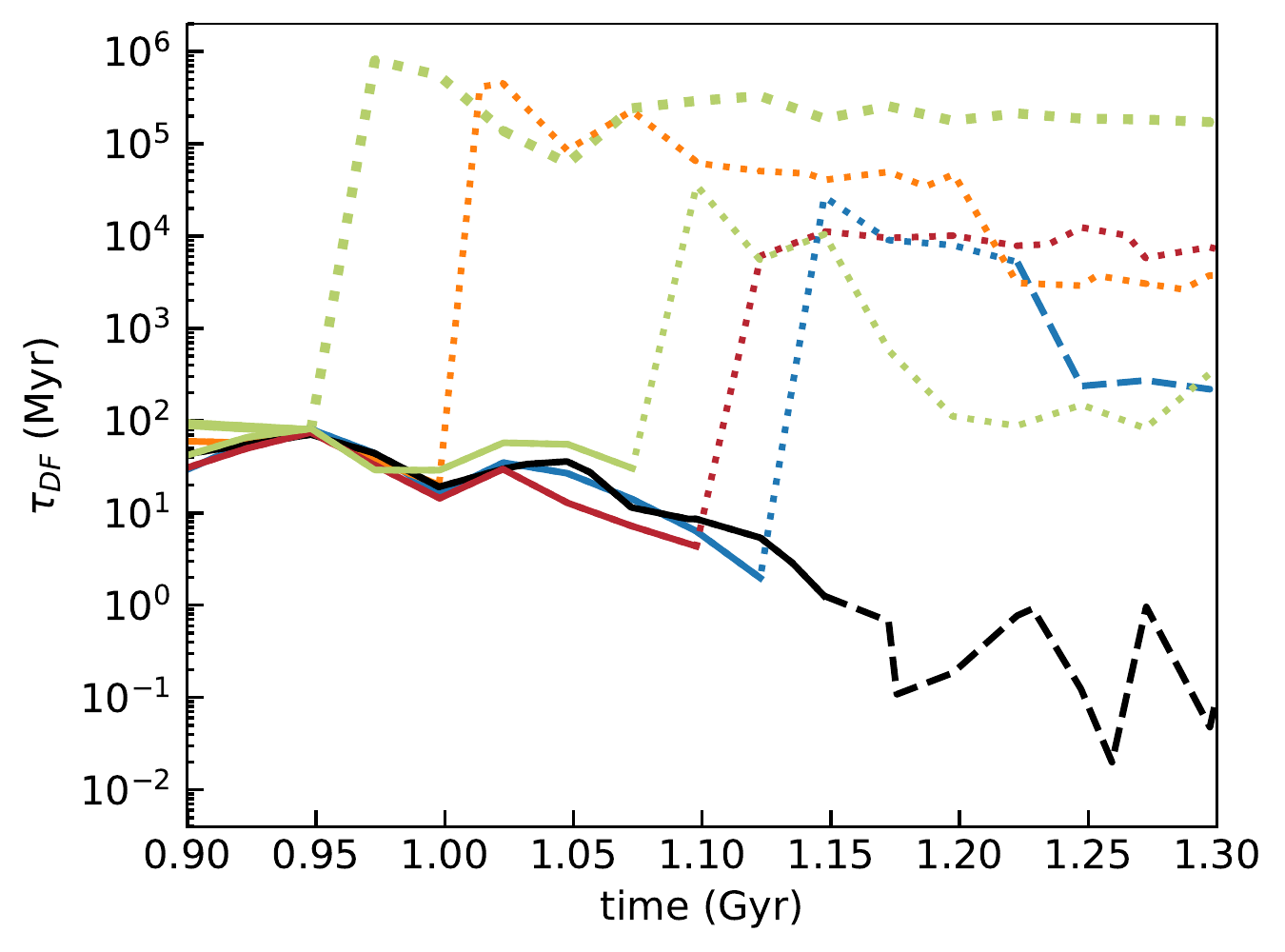}
\caption{Top panel: distance of the BH originally in the satellite galaxy from the center of the main galaxy as a function of time for a subset of the simulations listed in Table~\ref{table:CosmoSims}, as noted in the inset. If the BH in the satellite galaxy merges with the BH of the central galaxy, we show its subsequent evolution with dashed lines. Bottom panel: sinking time $\tau_{\rm DF}$, computed using Eq.~\eqref{eq:taudynamical friction}, as a function of time. We show the different phases: when the BH is still surrounded by material (solid line), when the BH is naked (dotted line; note the rapid increase in the sinking time because of the drop in $M_s$ in Eq.~\ref{eq:taudynamical friction}) and when the BH has merged with the BH of the central galaxy (dashed line).}
\label{fig:BHinHalo}
\end{figure}

\subsection{Formation of a black hole binary in a high-redshift galaxy merger}
\label{section:FormationOfABlackHoleBinaryInAHighZMerger}

We now focus on the same satellite galaxy, and follow the dynamical evolution of its BH during and after its host infalls into the halo of the larger galaxy. It is typically after this kind of event, when the galaxy remnant has settled and the massive BHs have sunk to the center of the potential well, that massive BH binaries form.

We show in Fig.~\ref{fig:BHinHalo} $\tau_{\rm DF}$ as a function of time, for all the simulations (bottom panel). We see that, initially, when the BH is still embedded in the satellite galaxy (solid line), its dynamics is the same for all simulations: the large scale dynamics is independent of the subgrid model we use. However, what happens following the disruption of the satellite galaxy (dotted line)  differs significantly from one simulation to the other: in some cases, the satellite BH sinks toward the center and ``merges'' (we recall that BHs are allowed to merge when they are separated by less than $4\Delta x$ and the kinetic energy of the binary is lower than the gravitational energy, but the real merger happens below our resolution) with the central BH of the main galaxy (the subsequent evolution is shown as a dashed line), in other cases, the BH stalls hundreds of pc away from the center. We also show in the top panel of Fig.~\ref{fig:BHinHalo} the distance of the satellite BH to the central galaxy it is sinking in.

We first compare the simulations \texttt{HR\_PD\_BH1e5} - \texttt{HR\_nPD\_BH1e5}, and \texttt{HR\_PD\_BH1e4} - \texttt{HR\_nPD\_BH1e4}, which differ only by the use or not of our subgrid model for dynamical friction from stars and dark matter. Fig.~\ref{fig:BHinOwnGal} shows that the model does not help in keeping BHs in the center, as discussed in \S \ref{section:InfallOfASeedBlackHoleInItsOwnGalaxy}: the galaxy is so chaotic that BHs wander no matter the implementation. When the galaxy is more settled, however, as it is  the case when the satellite BH falls into the main galaxy, we see the effects of our model (see Fig.~\ref{fig:BHinHalo}). When our prescription is used, the BH remains closer to the center; nonetheless the BHs do not merge as would happen if the BHs were artificially repositioned at the center of mass of the halo, as is sometimes done in cosmological simulations~\citep[e.g.][]{Vogelsbergeretal13, Schayeetal15}.

We now focus on simulations with $10^4 \Msun$ seeds. After its stellar and gaseous envelope has been dispersed (dotted line), the BH should take 1-100 Gyr to sink toward the center of the galaxy, and indeed, it stalls at $\sim$ hundreds of pc. This is in agreement with our understanding of dynamical friction: it is a very long process if the mass of the BH is low. The presence of a nuclear star cluster could speed-up the process \citep{2017MNRAS.469..295B}, increasing the mass experiencing dynamical friction, but due to our limited resolution, such compact structures of typical size of a few pc to $\sim$ ten of pc are not captured here~\citep{Georgievetal16}, and the envelope of the BH is rapidly stripped (dotted line). In the medium resolution case (\texttt{MR\_PD\_BH1e4}) the BH in the larger galaxy has also been scattered, similarly to what happened for the case studied in section \ref{section:InfallOfASeedBlackHoleInItsOwnGalaxy}. Accidentally, the two BHs merge while they are both off-center and  the remnant of this merger remains hundreds of pc away from the center. If we admit that this merger is physical, it is interesting to note that mergers of light seeds BHs are possible, though the dynamics is highly erratic. Multiple BHs in galaxies, each inherited from a different merger, are generically expected 
\citep[e.g.][]{1994MNRAS.271..317G,2002ApJ...571...30S,2005MNRAS.358..913V,Bonetti18-triplets,2018ApJ...857L..22T}.

Finally, we compare \texttt{HR\_PD\_BH1e5} and \texttt{HR\_PD\_BH1e4} which differ only by the seed mass of the BH. In \texttt{HR\_PD\_BH1e5}, the BH being more massive, it remains surrounded by a dense stellar concentration which does not disrupt (no dotted line), increasing even more the effective \Ms and resulting in a smooth  decay to the center of the main galaxy and a BH merger.

These experiments makes us believe that $<10^4\Msun$ seed BHs are less likely to contribute to the merging population observable by LISA than larger mass seed BHs. This does not exclude that these low-mass BHs may eventually sink in the center of galaxies and contribute to the massive BH population, but the presence of a dense stellar cluster or of bound gas on scales not resolved in this study, which would make the effective \Ms larger, appears to be crucial  \citep[e.g.][]{Callegari_09}.
Off-center mergers, happening by chance, as in \texttt{MR\_nGB\_PD\_BH1e4} could also contribute. 

\section{Conclusions}
We present a model to correct the dynamics of BHs in the {\sc ramses} code, which is currently the only code to include a physically motivated model for dynamical friction onto BHs from gas, stars and dark matter. We use this model in a suite of cosmological simulations to understand the dynamics of seed BHs ($10^4-10^5 \Msun$) in high-redshift galaxies and during galaxy mergers. We summarize our findings below:
\begin{itemize}
\item dynamical friction from stars has generically a more stabilizing effect than dynamical friction from gas, which can shock and is subject to inflows and outflows. In high-redshift galaxies, however, the stellar distribution is irregular and does not necessarily provide a smooth distribution within which BHs can decay undisturbed.  The presence of satellite galaxies can also perturb the orbit of a BH.
\item From the results of our best resolution cosmological simulation, BHs with masses of the order of $10^4 \Msun$ are subject to the fluctuations of the underlying stellar gravitational potential, which leads to a random walk-type of trajectory. This appears to be unique of 
a high-$z$ environment in which sub-structures undergo rapid evolution. If BHs were to be seeded in nuclear star clusters, or had masses of $10^5 \Msun$ or higher, they would be well stabilized galaxy centres. 
\item Similarly, following galaxy mergers, if the mass of BHs in satellite galaxies is $ \sim 10^4 \Msun$, it is unlikely that they participate in the merging population, although off-center mergers can occur fortuitously. If seed BHs have larger masses, $\sim 10^5 \Msun$, or they are embedded in dense bound stellar or gaseous envelopes, they can smoothly reach the center of the larger galaxy and merge with the companion BH.
\end{itemize}

\section*{Acknowledgments}
MV and HP acknowledge support from the European Research Council (Project no. 614199, `BLACK'). 
 HP acknowledges support from the COST Association (CA16104 Gravitational waves, black holes and fundamental physics), and thanks the University of Milano Bicocca for hosting him. This work was granted access to the HPC resources of under the allocations A0020406955 and A0040406955 made by GENCI. This work has made use of the Horizon Cluster hosted by the Institut d'Astrophysique de Paris; we thank Stephane Rouberol for running smoothly this cluster for us. Finally, we acknowledge Darren Croton for his useful comments.

\bsp	
\label{lastpage}
\end{document}